\documentclass{aastex62}

\graphicspath{{./}{figures/}}

\received{January 1, 2018}
\revised{January 7, 2018}
\accepted{\today}
\submitjournal{ApJ}

\shorttitle{Globular Cluster Systems and X-Ray Atmospheres in Galaxies}
\shortauthors{G. Harris et al.}
\begin{document}

\title{Globular Cluster Systems and X-Ray Atmospheres in Galaxies  \footnote{Released on January, 8th, 2018}}

\correspondingauthor{Gretchen L. H. Harris}
\email{glharris@uwaterloo.ca}

\author[0000-0002-9451-148X]{Gretchen L. H. Harris}
\affil{
Department of Physics and Astronomy, Waterloo Institute for Astrophysics\\
University of Waterloo \\
Waterloo, ON N2L 3G1, Canada}

\author{Iu. V. Babyk}
\affil{
Department of Physics and Astronomy, Waterloo Institute for Astrophysics\\
University of Waterloo \\
Waterloo, ON N2L 3G1, Canada}
\affil{
Main Astronomical Observatory of the National Academy of Sciences of Ukraine\\
Kyiv, 03143, Ukraine}
%\nocollaboration

\author{William E. Harris}
\affil{Department of Physics and Astronomy\\
McMaster University\\
Hamilton, ON L8S 4M1, Canada}
%\nocollaboration

\author{B.R. McNamara}
\affil{
Department of Physics and Astronomy, Waterloo Institute for Astrophysics\\
University of Waterloo \\
Waterloo, ON N2L 3G1, Canada}
%\nocollaboration

\begin{abstract}

We compare the empirical relationships between the mass of a galaxy's globular system $M_{GCS}$, the gas mass in the hot X-ray atmosphere $M_X$ within a fiducial radius of $5r_e$, the total gravitational mass $M_{grav}$ within $5 r_e$, and lastly the total halo mass $M_h$ calibrated from weak lensing.  We use a sample of 45 early-type galaxies (ETGs) for which both GCS and X-ray data are available; all the galaxies in our sample are relatively high-mass ones with $M_h > 10^{12} M_{\odot}$.  We  find that $M_X \propto M_h^{1.0}$, similar to the previously known scaling relation $M_{GCS} \propto M_h^{1.0}$. Both components scale much more steeply than the more well known dependence of total stellar mass $M_{\star} \propto M_h^{0.3}$ for luminous galaxies. These results strengthen previous suggestions that feedback had little effect on formation of the globular cluster system. 
The current data are also used to measure the relative mass fractions of baryonic matter and dark matter (DM) within $5 r_e$.  We find a strikingly uniform mean of 
$\langle f_{DM} \rangle = 0.83$ with few outliers and an rms scatter of $\pm 0.07$.  This result is in good agreement with two recent suites of hydrodynamic galaxy formation models.

\end{abstract}

%% Keywords should appear after the \end{abstract} command. 
%% See the online documentation for the full list of available subject
%% keywords and the rules for their use.

%% From the front matter, we move on to the body of the paper.
%% Sections are demarcated by \section and \subsection, respectively.
%% Observe the use of the LaTeX \label
%% command after the \subsection to give a symbolic KEY to the
%% subsection for cross-referencing in a \ref command.
%% You can use LaTeX's \ref and \label commands to keep track of
%% cross-references to sections, equations, tables, and figures.
%% That way, if you change the order of any elements, LaTeX will
%% automatically renumber them.
%%
%% We recommend that authors also use the natbib \citep
%% and \citet commands to identify citations.  The citations are
%% tied to the reference list via symbolic KEYs. The KEY corresponds
%% to the KEY in the \bibitem in the reference list below. 

\section{Introduction}\label{sec:intro}
If galaxies are formed through hierarchical growth by the merging of halos, 
%through gravitational instability,
then their global properties should correlate well with overall halo mass $M_h$, which is dominated by the dark-matter component.  But the reality is not that simple.  Starting from  early epochs, feedback mechanisms such as supernovae, stellar winds, and AGNs heat the gas in these halos, inhibiting further star formation and even driving the gas outward.  The relative strengths of these different mechanisms depend fairly sensitively on halo mass.  As is now well known, the ratio of stellar to halo mass (SHMR) is a very nonlinear function of $M_h$, reaching a peak near $M_h \simeq 10^{12} M_{\odot}$ and falling dramatically towards either lower- or higher-mass galaxies \citep[e.g.][among many others]{behroozi_etal2013, moster_etal2013, hudson_etal2015}.   

Galaxies in the high-mass regime $M_h \gtrsim 10^{12} M_{\odot}$ also typically hold two other interesting halo components:  substantial amounts of diffuse, hot X-ray gas; and large populations of globular clusters (GCs) also occupying the halo and galactic bulge.  The formation of the GCs in particular dates back to the very earliest stages of star formation \citep[e.g.][]{leaman_etal2013, forbes_etal2018a, brown_etal2018, choksi_etal2018, kruijssen_etal2019}.  The total numbers of GCs in a galaxy, and by extension the total mass $M_{GCS}$ in the entire system of GCs, have been found empirically to increase in almost direct proportion to $M_h$ over 5 orders of magnitude \citep[e.g.][]{blakeslee1997, spitler_forbes2009, hhh2015, hbh2017}, behaving unlike any other stellar population.

In this paper we directly compare recently compiled data for the total mass in globular clusters ($M_{GCS}$) with the observed global properties of the hot gaseous atmosphere in the same galaxies, including the gas mass $M_X$, temperature $T_X$, and luminosity $L_X$.  Whereas the GCS is a relic of the very earliest star formation, the development of a reservoir of hot gas is expected to take place over a longer period, during the main epochs of star formation and during intense periods of feedback (see below). We therefore do not expect any direct causal link between these two halo components.  Instead, the purpose of these comparisons is to see how each one correlates with the more fundamental halo mass, and give us a direct indicator of how strongly the overall feedback mechanisms have affected $M_{GCS}, M_X$, and the stellar mass $M_{\star}$ itself.

Recently \cite{james_etal2019} have discussed the relation between $M_{GCS}$ and galaxy total mass for ETGs, and their connections to luminosity $L_X$.  The aims of our paper are somewhat different: the observed correlations among 
the masses $M_{GCS}, M_h, M_X$ as well as $T_X$ and $L_X$ are presented more directly, and we work with a different sample of galaxies as well as a different calibration for the halo masses.  Although the GC systems and the hot gas 
content of galaxies are not usually intercompared it is also worth noting that, for a few large ellipticals, the radial distributions
of the X-ray gas and the GCs have been examined \citep[][and references cited there]{forbes_etal2012}, with the interesting
conclusion that the specific energy of the GC system is much lower than that of the hot gas.

The outline of this paper is as follows.  We provide background on the data for $M_{GCS}$, $M_h$, and $M_X$ in Section 2 and discuss our results in Sections 3, 4, and 5.  The data are all drawn from previous work.  Here, we select galaxies for which the global features of the GCS and the X-ray halo have both been measured in recent studies.  
We assume a distance scale $H_0 = 70$ km s$^{-1}$ Mpc$^{-1}$ for all the data.
 
\section{The Data}\label{sec:data}

\subsection{Globular Cluster System Mass}
The GCS catalogue of \citet{hha2013} (hereafter HHA) assembled data for 422 galaxies with published measurements of their globular cluster systems.  The sample consisted of 321 early-type galaxies (ETGs, including elliptical and lenticular types) and 81 S/Irr types. 
The main data for these galaxies included the estimated total globular population, $N_{GC}$, and their total mass $M_{GCS}$.  HHA discussed the method of determining $M_{GCS}$ from $N_{GC}$ and this was slightly revised in the recalibration of mass-to-light ratio for globular clusters by \citet{hbh2017} (hereafter HBH).  The values for $M_{GCS}$ used here are based on the HBH calibration.  

One factor to consider for the definition of $M_{GCS}$ is that it includes both metal-poor and metal-rich GCs (with a rough dividing line at [Fe/H] $\sim -1$), but the relative numbers of blue (metal-poor) and red (metal-rich) clusters vary strongly with host galaxy mass (see HHA, \citet{hhh2015} hereafter HHH2).  However, in practice the dependence of $M_{GCS}$ on $M_h$ is driven primarily by the metal-poor GCs, since almost 80\% of all GCs summed over all galaxies belong to the metal-poor category \citep{harris2016}, consistent with their very early formation in small halos \citep{choksi_gnedin2019}.

\subsection{Halo Mass}

The dark-matter-dominated halo is the potential well within which all the galaxy's baryonic material
accumulates and evolves.  But the total halo mass is a difficult quantity to measure 
particularly because it extends to 
such large radius beyond the baryonic components, and the $M_h$ estimate for any single galaxy will inevitably have significant uncertainty. 
The values for $M_h$ used here are also drawn from HBH and are based on the work of 
\citet{hudson_etal2015} who used weak lensing to calibrate $M_h$ versus near-IR luminosity $L_K$. They determined a homogeneous SHMR built from a single mass determination method over nearly five orders of magnitude in galaxy halo mass.  \citet{hhh2014}(hereafter HHH1) used V and K band luminosities and the prescriptions of \citet{bell_etal2003} to determine mass-to-light ratios for the galaxies in the GCS catalogue of HHA.  Then SHMR from weak lensing was used to calculate halo mass, defined as $M_h = M_{200} + M_*$, where $M_{200}$ is defined as the dark matter halo mass within a radius in which the mean density is 200 times the critical density.  
%%In summary, $M_h$ represents the total gravitating material in the galaxy.  

 The correlation of SHMR with $M_h$ used here is built entirely on weak lensing, but it is worth stating that calibrations of SHMR constructed from other methods such as satellite dynamics, abundance matching, or halo occupation index, give very similar results for low-redshift systems 
\citep[e.g.][among others]{behroozi_etal2013, leauthaud_etal2012, vanUitert_etal2016, wechsler_tinker2018}, so our results to be discussed below do not depend strongly on one particular approach to defining the SHMR.  This mass ratio has a maximum value of a few percent for mid-range galaxies near 
$M_h \sim 10^{12} M_{\odot}$ and declines steeply towards both dwarfs and giants, a result of the relative effects of feedback on the efficiency of star formation; for modelling discussions of feedback (reionization, stellar winds, supernovae, AGNs, infall heating)
and its effects, see, e.g., \citet{behroozi_etal2013, mitchell_etal2016, agertz_kravtsov2016, wechsler_tinker2018}.

\subsection{X-ray Data and Hot Gas Mass}

\begin{table*}[t!]
\caption{The derived quantities of our sample.}\label{tab1}
\centering
\begin{tabular}{lcccccccccccc}
\hline
 && \\
Name & $5r_e$ & $\log(M_{GCS}$)& $\log(M_{h}$) & $\log(M_{*}$)& T$_{X}$ & L$_{X}$ & $\log(M_{X}$)& $\log(M_{grav})$ & $f(DM)$ \\
   & kpc  & M$_{\odot}$ & M$_{\odot}$ & M$_{\odot}$  &   keV  & 10$^{40}$erg/s & M$_{\odot}$& M$_{\odot}$ & \\
        & (2) & (3) & (4) & (5) & (6) &(7) & (8) & (9) & (10)\\
&& \\
\hline
&&\\
IC1459  & 35$\pm$6&8.82$\pm$0.13&13.6&11.4 & 0.70$\pm$0.01 &2.41$\pm$0.06&  10.10 & 12.08 & 0.78$\pm$0.15\\
IC4296  & 77$\pm$9&9.37$\pm$0.02&14.2 &11.6 & 0.94$\pm$0.01& 18.11$\pm$0.42& 10.66 & 12.74 & 0.92$\pm$0.11\\
NGC708  & 110$\pm$14&9.18$\pm$0.08&13.3 & 11.2 & 1.56$\pm$0.01& 193.0$\pm$10.9 & 11.64 & 12.69 & 0.88$\pm$0.25 \\
NGC720  & 32$\pm$3&8.32$\pm$0.11&13.3 &11.2 & 0.62$\pm$0.01& 4.11$\pm$0.06 & 10.78 & 11.97 & 0.77$\pm$0.14\\
NGC821  & 30$\pm$7&7.97$\pm$0.06&12.6 &10.9  & 0.20$\pm$0.08 & 0.10$\pm$0.03 & 9.50 & 11.50 & 0.74$\pm$0.11\\
NGC1023 & 35$\pm$4&8.14$\pm$0.03&12.5&10.9 & 0.20$\pm$0.09 & 0.21$\pm$0.02 & 9.60 & 11.52 & 0.75$\pm$0.12\\
NGC1316 & 58$\pm$7&8.57$\pm$0.18&14.2&11.6 &0.75$\pm$0.01 & 6.47$\pm$0.09 & 10.96 & 12.33 & 0.77$\pm$0.15\\
NGC1332 & 30$\pm$4&8.47$\pm$0.18&13.0&11.1 & 0.70$\pm$0.03 & 2.16$\pm$0.04 & 9.51 & 12.28 & 0.93$\pm$0.11\\
NGC1399 & 45$\pm$5&9.28$\pm$0.05&13.5&11.3 & 1.26$\pm$0.03 & 24.96$\pm$1.12 & 10.67 & 12.48 & 0.92$\pm$0.12 \\
NGC1404 & 45$\pm$5&8.36$\pm$0.07&13.0 &11.1 & 0.67$\pm$0.04 & 26.9$\pm$10.2 & 10.77 & 12.15 & 0.87$\pm$0.12\\
NGC1407 & 44$\pm$5&9.4$\pm$0.11&13.9 &11.5 & 1.02$\pm$0.02 & 5.77$\pm$0.13 & 10.98 & 12.33 & 0.81$\pm$0.14\\
NGC1600 & 97$\pm$10&9.04$\pm$0.07&14.4&11.7 & 1.24$\pm$0.02 & 17.14$\pm$0.40 & 11.62 & 12.87 & 0.88$\pm$0.11\\
NGC1700 & 45$\pm$6&8.63$\pm$0.08&13.4&11.3 & 0.51$\pm$0.02 & 7.81$\pm$0.18 & 10.99 & 12.11 & 0.77$\pm$0.13\\
NGC2434 & 25$\pm$3&7.61$\pm$0.11&12.3 &10.7 & 0.59$\pm$0.03 & 0.79$\pm$0.04 & 10.38 & 11.70 & 0.85$\pm$0.13\\  
NGC2768 & 31$\pm$4&8.36$\pm$0.12&13.2&11.2 & 0.35$\pm$0.02 & 0.74$\pm$0.02 & 10.73 & 11.61 & 0.48$\pm$0.25\\
NGC3379 & 29$\pm$5&7.77$\pm$0.08&12.4 &10.8 & 0.24$\pm$0.08 & 0.18$\pm$0.02 & 9.30 & 11.51 & 0.80$\pm$0.17\\
NGC3384 & 23$\pm$3&7.49$\pm$0.1&12.2&10.6 & 0.31$\pm$0.04 & 0.09$\pm$0.01& 9.11 & 11.48 & 0.86$\pm$0.13\\
NGC3557 & 60$\pm$7&8.61$\pm$0.23&14.3 &11.7 & 0.43$\pm$0.10 & 4.91$\pm$0.67& 11.61 & 12.85 & 0.87$\pm$0.11\\
NGC3585 & 35$\pm$4&7.96$\pm$0.12&13.1 &11.2 & 0.32$\pm$0.07 & 0.28$\pm$0.03& 9.05 & 11.71 & 0.69$\pm$0.13\\
NGC3607 & 28$\pm$3&8.25$\pm$0.12&12.9 &11.0 & 0.59$\pm$0.11 & 0.73$\pm$0.03 & 9.06 & 11.72 & 0.81$\pm$0.17\\
NGC3923 & 50$\pm$6&8.96$\pm$0.04&13.6 &11.3 & 0.58$\pm$0.01& 5.02$\pm$0.06& 10.62  & 12.18 & 0.90$\pm$0.13\\
NGC4073 & 104$\pm$16&9.5$\pm$0.03&14.5 &11.8 & 1.88$\pm$0.02&225.2$\pm$21.6 & 11.47 & 12.91 & 0.89$\pm$0.13\\
NGC4203 & 21$\pm$3&7.63$\pm$0.2&12.2 &10.6 & 0.28$\pm$0.03 & 0.44$\pm$0.05 & 9.05 & 11.62 & 0.90$\pm$0.14\\  
NGC4261 & 45$\pm$4&8.6$\pm$0.08&13.5 &11.3 & 0.80$\pm$0.01 & 6.61$\pm$0.08 & 10.36 & 12.33 & 0.90$\pm$0.14\\
NGC4278 & 25$\pm$3&8.48$\pm$0.1&12.4 &10.8  &  0.33$\pm$0.02 & 0.35$\pm$0.01 & 9.04 & 11.71 & 0.87$\pm$0.16\\   
NGC4365 & 28$\pm$2&9.03$\pm$0.07&13.6 &11.4 & 0.44$\pm$0.02& 0.41$\pm$0.02 & 10.33 & 11.63 & 0.36$\pm$0.34 \\
NGC4374 & 33$\pm$3&9.15$\pm$0.11&13.5 &11.3 & 0.81$\pm$0.01& 5.18$\pm$0.05 & 10.56 & 12.15 & 0.83$\pm$0.14\\
NGC4382 & 34$\pm$3&8.56$\pm$0.07&13.4 &11.3 & 0.44$\pm$0.03& 1.13$\pm$0.03 & 10.20 & 11.63 & 0.50$\pm$0.23 \\
NGC4406 & 35$\pm$3&8.97$\pm$0.03&13.5 &11.3 & 0.88$\pm$0.01 & 10.28$\pm$0.24 & 11.09 & 12.09 & 0.74$\pm$0.12\\
NGC4472 & 36$\pm$3&9.39$\pm$0.05&14.0 &11.5 & 1.06$\pm$0.02 & 18.36$\pm$2.08 & 10.84 & 12.31 & 0.81$\pm$0.18\\
NGC4486 & 35$\pm$4&9.65$\pm$0.03&13.7 &11.4  & 1.85$\pm$0.02& 262.9$\pm$30.2 & 11.53 & 12.61 & 0.86$\pm$0.12\\
NGC4526 & 25$\pm$3&8.06$\pm$0.11&13.0 &11.1 & 0.37$\pm$0.02& 0.32$\pm$0.02 & 10.06 & 11.72 & 0.74$\pm$0.19 \\
NGC4552 & 24$\pm$2&8.44$\pm$0.07&12.7 &10.9 & 0.64$\pm$0.01& 2.48$\pm$0.04& 10.30 & 11.85 & 0.86$\pm$0.13\\
NGC4564 & 13$\pm$3&7.71$\pm$0.06&12.1&10.4 & 0.38$\pm$0.15& 0.09$\pm$0.02 & 9.05 & 11.72 & 0.95$\pm$0.13\\ 
NGC4621 & 24$\pm$3&8.36$\pm$0.16&12.7 &10.9 & 0.26$\pm$0.07& 0.02$\pm$0.004 & 10.11 & 11.51 & 0.71$\pm$0.20\\
NGC4636 & 34$\pm$3&9.09$\pm$0.01&12.8 &10.9 & 0.75$\pm$0.03 & 19.68$\pm$1.09& 11.06 & 12.04 & 0.82$\pm$0.12\\
NGC4649 & 41$\pm$4&9.13$\pm$0.05&13.7 &11.4 & 0.94$\pm$0.03 & 12.13$\pm$1.06 & 10.62 & 12.31 & 0.86$\pm$0.13\\
NGC4697 & 35$\pm$3&7.81$\pm$0.12&12.5 &10.9 &0.31$\pm$0.01 & 0.64$\pm$0.03 & 11.08 & 11.51 & 0.38$\pm$0.18\\
NGC5018 & 39$\pm$4&8.52$\pm$0.11&13.3 &11.2 & 0.53$\pm$0.07& 1.52$\pm$0.10& 10.43 & 12.11 & 0.86$\pm$0.12\\
NGC5813 & 32$\pm$2&8.98$\pm$0.06&13.5 &11.3 & 0.71$\pm$0.02 & 43.9$\pm$10.1 & 11.33 & 11.85 & 0.42$\pm$0.15\\
NGC5846 & 34$\pm$6&9.17$\pm$0.1&13.4 &11.3 & 0.79$\pm$0.03& 15.72$\pm$0.11 & 11.01 & 12.07 & 0.74$\pm$0.14\\
NGC5866 & 26$\pm$3&8.01$\pm$0.08&12.5 &10.8  & 0.41$\pm$0.08& 0.33$\pm$0.02 & 10.42 & 11.52 & 0.73$\pm$0.12\\
NGC6482 & 37$\pm$4&8.59$\pm$0.03&13.4 &11.3 & 0.82$\pm$0.01 & 63.39$\pm$0.88 & 11.21 & 12.21 & 0.78$\pm$0.12\\
NGC6861 & 41$\pm$4&8.73$\pm$0.12&12.8 &11.0 & 1.24$\pm$0.03 & 6.24$\pm$0.14 & 10.52 & 12.51 & 0.96$\pm$0.11\\
NGC7626 & 52$\pm$3&8.98$\pm$0.04&13.7 &11.4 & 0.93$\pm$0.02 & 7.19$\pm$0.33 & 10.95 & 12.62 & 0.92$\pm$0.12\\
\hline
\end{tabular}
\end{table*}

The X-ray data used here are from \cite{babyk_etal2018} (hereafter B18), who studied the scaling relations for 94 ETGs mostly within 100 Mpc.  To summarize briefly their analysis, first X-ray Chandra observations for galaxies with cleaned exposure times above 10 ks were selected and downloaded from the HEASARC\footnote{https://heasarc.gsfc.nasa.gov/} archive. Such a high exposure time eliminates large errors of parameters during the further spectral fitting.  The CIAO v.4.8 software package and CalDB v.4.7.1 were then used to extract  exposure- and background-corrected X-ray images in the 0.5-6.0 keV energy band. Point sources and other non-X-ray-gas features were
detected and then removed by applying the \texttt{wavedetect} routine. 

The X-ray spectra were extracted from a circular region within a fiducial radius of 5 effective radii, where
$r_e$ was derived from the optical Digitized Sky Survey. The multi-component spectral models \texttt{PHABS∗(APEC+PO+MEKAL+PO)} were applied to model each spectrum with the Xspec v.12.9.1 tool \citep{arn}. Such a combination of spectral parameters helps to define the contribution of unresolved low-mass X-ray binaries (LMXBs), active binaries, cataclysmic variables, and other stellar sources that contribute to the total X-ray emission. The \texttt{APEC} component models the thermal emission from the hot atmosphere. Power-law (\texttt{PO}) and a set of \texttt{MEKAL+PO} \citep{Mewe, Liedahl} components model the thermal and non-thermal X-ray emission from the LMXBs and other stellar sources, while the \texttt{PHABS} component models the photoelectric absorption. The spectral fitting then gave an average temperature and luminosity for each galaxy. These results are in good agreement with previous analyses \citep{boroson, su, gould}.

The assumption of hydrostatic equilibrium was used to calculate both $M_X$ and $M_{grav}$. The extracted surface brightness profiles from the X-ray images were fitted with a single $\beta$-model \citep{caval} as 
\begin{equation}
    S(r) = S_0\left( 1+\left(\frac{r}{r_c}\right)^2\right)^{-3\beta+0.5} + C,
\end{equation}
where $S(r)$ is the brightness profile as a function of radius, while $r$, $S_0$, $r_c$, $\beta$ and $C$ are free parameters in the fit. The typical slope $\beta$ is 0.4-0.5.  The gas density profile is then 
\begin{equation}
    \rho_g(r) = \rho_0\left(1+\left(\frac{r}{r_c}\right)^2\right)^{-3\beta/2},
\end{equation}
where $\rho_0$ = 2.21$\mu m_p n_0$ is the central gas density and $n_0$ is the central concentration that can be found from the emissivity \citep[see][and B18 for more details]{ettori}. The hot-gas mass was calculated by integrating the gas density profiles within $r$, as 
\begin{equation}
    M_X(r) = 4 \pi \rho_0 \int_0^r r^2 \left(1+\left(\frac{r}{r_c}\right)^2\right)^{-3\beta/2} dr.
\end{equation}

%%The $M_X$ values listed here in Table 1 are the correctly calculated gas masses
%%enclosed within $5r_e$.) }

The total gravitational mass within $r$ is calculated from 
\begin{equation}
    M_{grav}(r) = -\frac{k T r}{G\mu m_p} \left(\frac{d \ln{\rho_g}}{d\ln{r}}\right).
\end{equation}

(NB:  The values for ${M_{X}}$ in B18 (their ${M_g}$) are incorrect and an Erratum is being prepared.  Our values in Table 1 are the corrected ones and do not correspond with those in the original paper.) 

As noted above, the gas component as we use it here refers to the hot gas within the galaxy, 
inside the fiducial radius of $5r_e$.  It does not include
any gaseous ICM (IntraCluster Medium) that would (if present) be distributed across
much larger scales.

$M_{grav}$ as determined by B18 is essentially a `hydrostatic mass' and is a measure of the total gravitating mass within a given outer radius, which we adopt here as $5 r_e$. Empirically, $5r_e$ is sufficiently large to enclose almost all the currently measurable hot gas in the galaxies (see B18), and as will be seen below, it is a useful fiducial radius for various comparisons with theory.  However, since $r_e \ll r_{200}$ for large galaxies, and since dark matter becomes increasingly dominant at larger radii, the halo mass $M_h$ as described in the previous section is generally much larger than $M_{grav}$.

Uncertainties remain in $M_{grav}$ that are hard to assess particularly in the
low-temperature regime $T_X \lesssim$ 0.5 keV where the Chandra instruments are less effective and $L_X$ is low (see also the discussion below).  For these lower-mass galaxies measurements of $r_e$ itself also show increased
scatter (see B18).  The increased scatter at the low$-T_X$ end of the correlations should be viewed with these unavoidable issues in mind.

In B18, the scaling relations derived for $T_X, L_X, M_X, M_{grav}$ were found to be a bit different 
from those for larger-scale clusters or groups of galaxies.   These scaling relations are also very different from the ones expected from self-similar scaling, strongly suggesting that the gas has been affected by extra heat sources, particularly AGN feedback.  A detailed theoretical modelling of the mechanisms is still needed.

\begin{figure}
    \centering
    \includegraphics[width=1.0\textwidth]{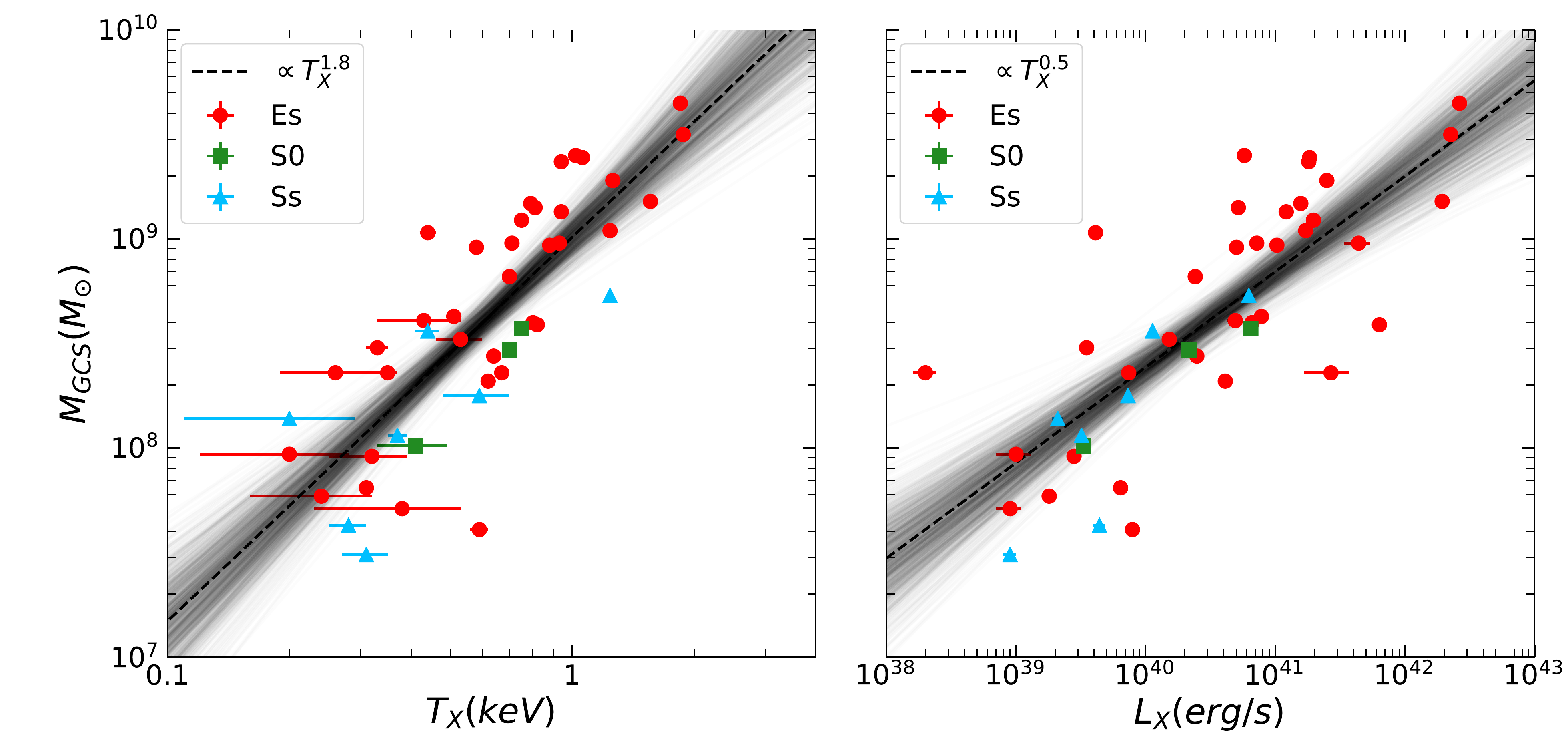}
    \caption{The mass of a galaxy's globular cluster system, $M_{GCS}$ ($M_{\odot}$), is plotted versus the temperature $T_X$ (keV) and luminosity  $L_X$ (erg/s) of the X-ray halo surrounding the galaxy.  E galaxies are red circles, S0 green squares and S blue triangles.  The dashed lines give the best-fit correlations, while the shaded areas represent best-fit slope uncertainties.}
    \label{figure1}
\end{figure}

\section{Correlations}

Between the X-ray galaxy sample of B18 and the GCS sample of HBH, there are 45 galaxies in common, all of which have $M_h > 10^{12} M_{\odot}$. GCs are found in all galaxies except the smallest dwarfs, but the presence of halo gas sufficiently hot to be detected in X-rays is restricted to these more massive systems.  Nevertheless, the overlap in the samples is still enough to cover 3 orders of magnitude in $M_h$ and thus to give enough leverage to estimate power-law scalings.  

The galaxies in this overlap sample and their properties are summarized in Table 1.  Here, $M_{GCS}, M_{\star}$, and $M_h$ are taken from \citet{hha2013, hbh2017}, and the other quantities $5r_e,T_X, L_X, M_X,$ and $M_{grav}$ from B18.  
The quoted individual uncertainties are from the source papers as well.  Uncertainties in $M_h$ and $M_{\star}$ are harder to gauge on an individual basis, but $M_h$ is expected to be uncertain by typically $\pm0.2$ dex in the mean from its weak lensing calibration
\citep[see][]{hhh2014, hudson_etal2015}.  The stellar mass $M_{\star}$ is determined from the measured galaxy luminosity $L_V$ or $L_K$ and an appropriate mass-to-light ratio; it is also likely to be uncertain to at least $\pm0.2$ dex \citep{hha2013}.

%We used the bivariate correlated error and intrinsic scatter (BCES) routine \citep{bces} to determine the form of the scaling relations. This routine performs a linear least-squares regression that minimizes the orthogonal distances of the datapoints to the best-fit relation. We also use 10,000 Monte Carlo bootstrap resamplings to determine the parameter uncertainties.
We used the likelihood-based method of \citet{Kelly:07} to determine the form of the scaling relations. This is one of the best regression methods for both improved confidence intervals and bias removal. This method is a Bayesian approach based on estimating a likelihood function and assumes the presence of intrinsic scatter in the parameters.  We used 15,000 iterations of a Markov Chain Monte Carlo to define the parameter uncertainties. 

We first show, in Figure 1, the correlations between $M_{GCS}$ and X-ray atmosphere temperature and luminosity.  Again, we emphasize that we do not expect any causal relationship between the two components, but both of them are useful indicators of the 
total gravitational potential of their host galaxy.  For $T_X$ we find $M_{GCS} \sim T_X^{1.8 \pm 0.2}$ with scatter increasing noticeably toward lower luminosity.   The correlation with $L_X$ exhibits more scatter over the entire range, but yields $M_{GCS} \sim L_X^{0.5 \pm0.06}$.  The $M_{GCS} - T_X$ correlation in particular suggests that the total GC system mass is a reasonable proxy for the halo gas mass, at least for galaxies massive enough to contain hot X-ray gas. 

In Figure 2, the correlations of $M_{GCS}$, $M_X$, and $M_{grav}$ with $M_h$ are shown. For a sample of 293 galaxies over the much larger mass range $M_h \simeq 10^{10} - 10^{15} M_{\odot}$,  \cite{hhh2015} found $M_{GCS} \sim M_h^{1.03\pm0.03}$, which is the line shown in the upper panel of Fig.~2. 
For the $M_{GCS}$ graph, the handful of most massive galaxies ($> 10^{14} M_{\odot}$) fall noticeably below the line defined by the smaller systems, but at least part of this offset is expected to be due to incompleteness:  
for these largest cD-type and BCG-type galaxies, the GC systems are extremely spatially extended and the current $N_{GC}$ values in the catalog are certainly underestimates \citep[see, e.g.][]{hha2013, harris_etal2017}.  We note, however, that the models of \cite{choksi_gnedin2019}
show a gradual downturn of $M_{GCS}$ versus $M_h$ for $M_h \gtrsim 10^{13} M_{\odot}$, which is at least roughly similar to what we see in Fig.~2.  In current models a significant fraction of the GCS for these highest-mass galaxies is acquired by accretion of small galaxies, which contribute much dark matter but whose 
GCs have lower-than-average mass \citep[see the discussion of][]{choksi_gnedin2019}.  The other scalings were calculated for $M_h < 10^{14} M_{\odot}$.

In the middle panel of Fig.~2 the best-fit line 
$M_X \sim M_h^{1.0 \pm 0.2}$ derived from our data is shown. 
In brief, both $M_X$ and $M_{GCS}$ scale almost in direct proportion
to the total gravitating mass of the host galaxy.
Though the \emph{temperature} and thus the luminosity of the X-ray gas increase steeply with galaxy mass
(the results of progressively stronger heating from feedback), the total \emph{mass} in the X-ray gas reservoir appears to stay almost directly proportional to $M_h$.

%Figure 3 shows the same comparisons now as mass ratios of $M_{GCS}/M_h$, $M_X/M_h$, and %$M_{grav}/M_h$. 
Figure 3 shows, for completeness, the reverse correlations where now $M_{\star}, M_{grav}, M_X, M_h$ are each plotted versus $M_{GCS}$.  This graph is perhaps most useful to show the different degrees of scatter in each case (largest for $M_X$).
Lastly, in Figure \ref{figure4} we show a new mass ratio, hot X-ray gas mass to GCS mass, plotted against $M_h$.  The line shown in the graph has a slope of $-0.1$ consistent with the expected dependence from the slopes plotted in Fig.~2.  Table 2 lists the fitted slopes, zeropoints, and uncertainties for each correlation.

We also measure and list in Table~\ref{tab2} the intrinsic scatters of our scalings, showing how close the data distributions are to linearity. Unsurprisingly, the largest intrinsic scatter is found in the scalings that include X-ray gas measurements, i.e. $M_X$ and $M_{GCS}/M_X$ vs. $M_h$. Several factors may be contributing to the scatter, including departures from hydrostatic equilibrium \citep{Fabjan:11} and so-called baryon physics associated with assumed gas properties and uncertainties in the heating and cooling rates introduced by feedback that affect the atmosphere emission measure \citep{Fabjan:11}.

The $M_X$  values used here do not include any gas 
that might be present at radii outside $\sim 5 r_e$. Indeed, the amounts of such
sparse gas are not well known for most galaxies.  But particularly for the BCGs sitting
at the centers of rich clusters, an IntraCluster Medium (ICM) on much larger scales can be present 
\citep[e.g.][among many others]{gould}.  The great majority of galaxies in
our current list are not in this category, however, and many are relatively isolated.  For these the choice of
$5r_e$ as a fiducial limiting radius is large enough to include most of the gas 
hot enough to be measureable through X-rays (see B18).  
The X-ray surface brightness falls with radius, and  
current X-ray instrumentation, including Chandra, is incapable of detecting the X-ray emission from a dilute plasma at such very large radii.
%Therefore, the cutoff at $5r_e$  may underestimate the total atmospheric gas mass over all radii by up to a factor of two.
Nevertheless, the few BCGs in the list (such as M87, NGC 1399, NGC 708) do not themselves deviate
systematically from the correlations described above.
%ossible reason for their higher scatter may be the degree of regularity of our sample. The scatter can be significant in objects that do not follow hydrostatic equilibrium \citep{Fabjan:11}, in systems that depart from spherical symmetry \citep{Limousin:13}, or in very morphologically complex haloes \citep{Sereno:16}. It is well known that systems with a temperature of X-ray gas below 1-2 keV show significant deviations from hydrostatic equilibrium and their X-ray emission can not be explained by classical Bremsstrahlung scattering. \citet{Fabjan:11} showed that the degree of scatter depends on the baryonic physics as well. They also showed that the scatter in less spherical and irregular systems is increased with redshift. Since our sample consist of nearby systems ($z = 0$) we are not able to explore this effect. 
%Our sample includes both early- and late-type galaxies. The largest scatter is associated with the ellipticals which are mostly relaxed systems. 
%Another source of intrinsic scatter is that all our observables are time dependent and as a result, the slopes of any scalings can be time dependent too. But again, since our sample consists of nearby objects this can not explain our large scatter for $M_X$ vs. $M_h$ and $M_{GCS}/M_X$ vs. $M_h$ scalings. The most likely issue  is related to hydrostatic equlibrium or the Bremsstrahlung regime for the low-temperature gas.

\begin{figure}
    \centering
    \includegraphics[width=1.0\textwidth]{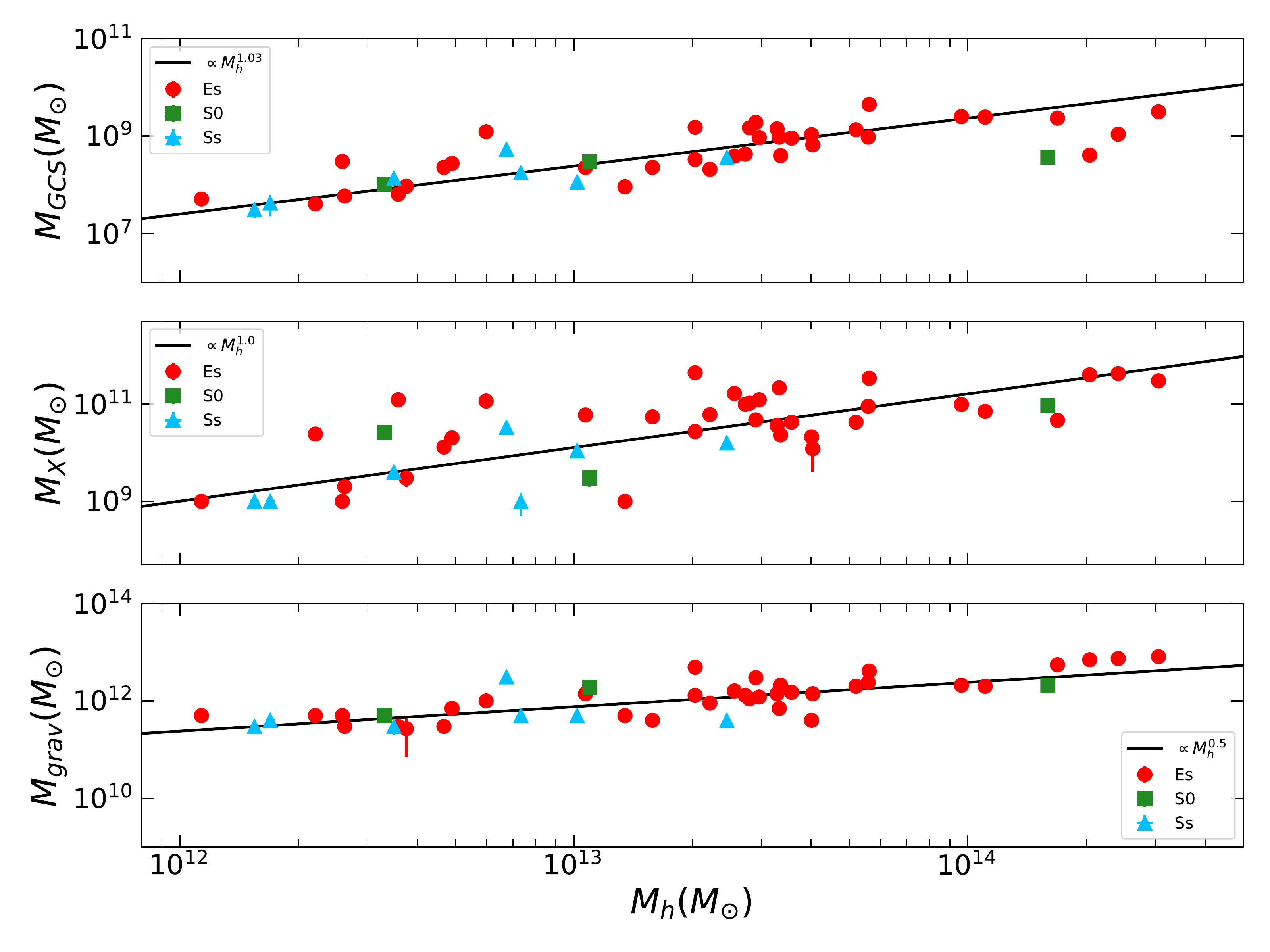}
    \caption{$M_{GCS}$  (upper panel), $M_X$ (middle panel), and $M_{grav}$ (lower panel) are plotted versus $M_h$  for the 45 galaxies in our overlap sample.  Symbols are as in Figure 1.  In the top panel, the dashed line shows $M_{GCS} \propto M_h^{1.03}$ as found by \citet{hhh2014, hhh2015} for a larger sample of galaxies. In the middle panel, the line shows the best-fit solution 
    $M_X \propto M_h^{1.0}$;  while in the lower panel, the line shows the best-fit $M_{grav} \propto M_h^{0.5}$.}
    \label{figure2}
\end{figure}

\begin{figure}
    \centering
    \includegraphics[width=1.0\textwidth]{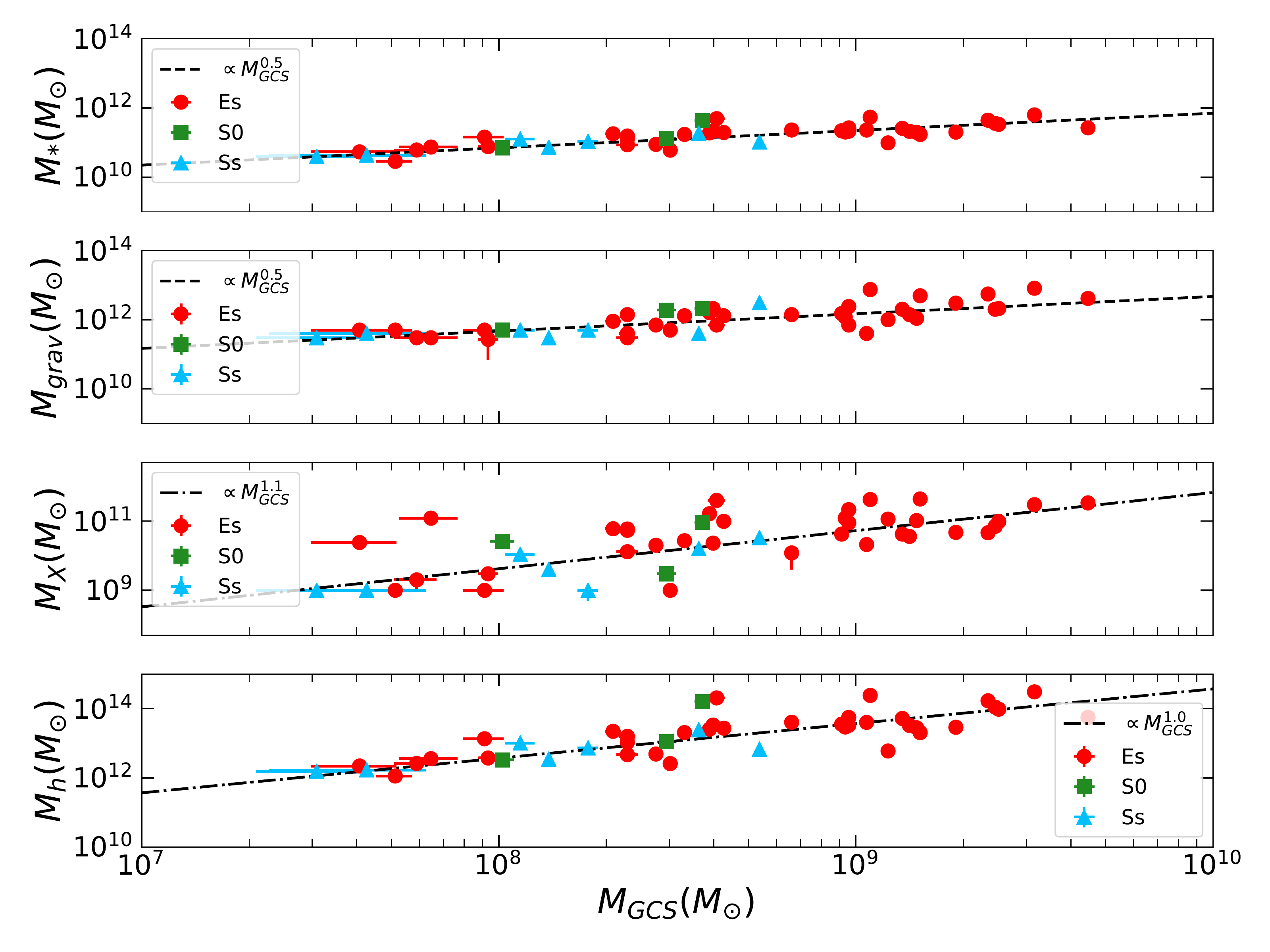}
    \vspace{-0.9cm}
    \caption{The four masses $M_{\star}, M_{grav}, M_X$ and $M_h$ here are plotted versus M$_{GCS}$. Symbols are as in Figure 1.}
    \vspace{0.5cm}
    \label{figure3}
\end{figure}

\section{Discussion}

The scalings for the GCS and X-ray components can also be compared with total \emph{stellar} mass $M_{\star}$.   For galaxies over the same mass range $M_h \gtrsim 10^{12} M_{\odot}$ as we are discussing here,  $M_{\star}$ is seen to increase as a distinctly shallower power of $M_h$, reflecting the increasing dominance of dark matter for more massive galaxies.  Discussions of various samples of galaxies at low redshift give power-law scalings in the range $M_{\star} \propto M_h^{0.25-0.5}$ \citep[e.g.][]{behroozi_etal2013, moster_etal2013, hudson_etal2015, kravtsov_etal2018, vanUitert_etal2016}.  These are galaxies falling on the high-mass side of the SHMR peak, and though different interpolation models have been used to fit the data along with various methods for determining halo mass (see the references cited above), a relatively simple dependence shows up that can be described to first order by a simple power-law scaling.

\begin{figure}
    \centering
    \includegraphics[width=0.8\textwidth]{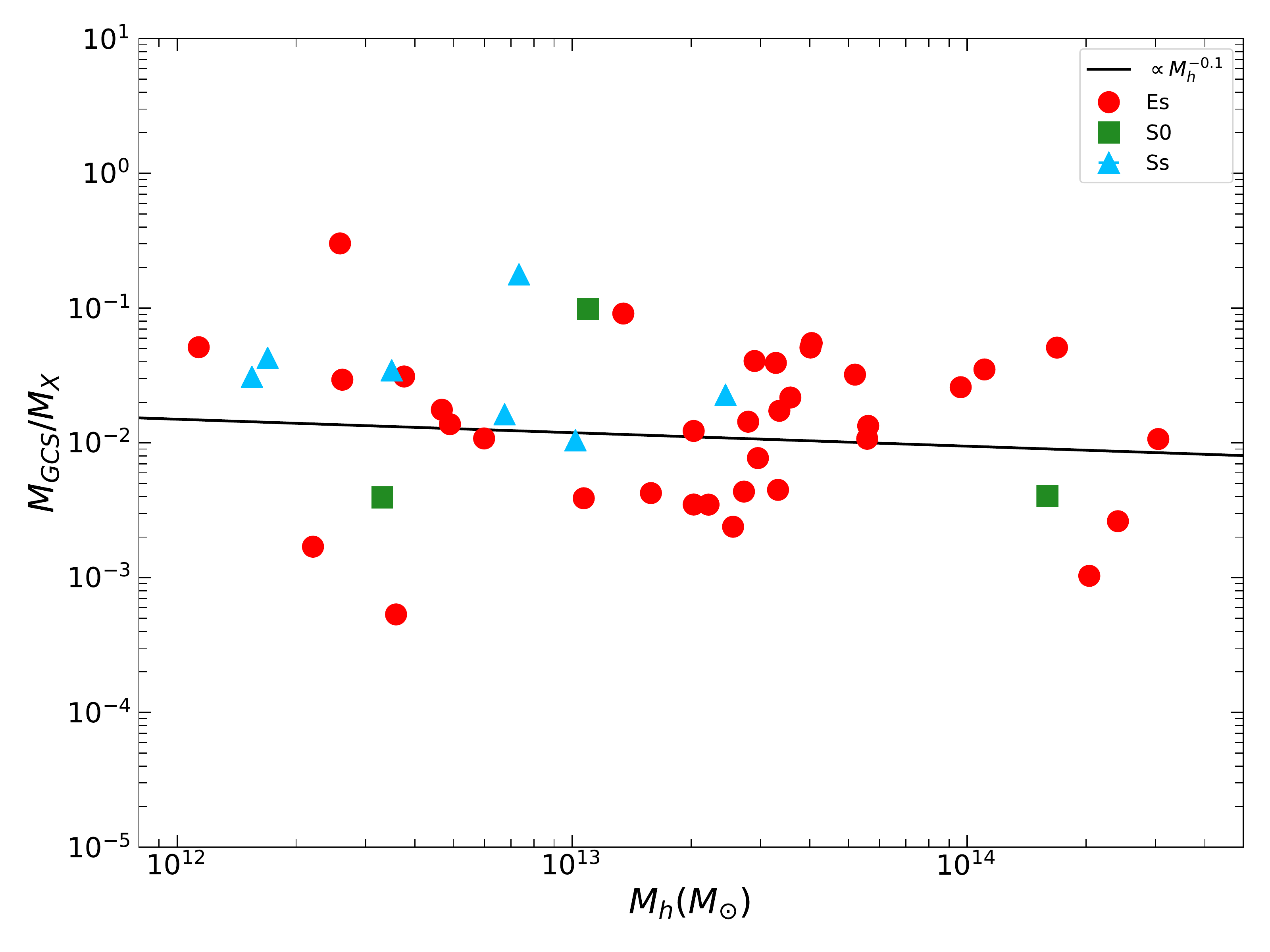}
    \caption{The ratio of  $M_{GCS} / M_X$ is plotted versus halo mass $M_h$. Symbols are as in Figure 1. The overplotted line is for $(M_{GCS} / M_X)  \propto M_h^{-0.1}$, consistent with the slopes in Figure 2. }
    \label{figure4}
\end{figure}

For comparison, \citet{james_etal2019} plot $M_{GCS}$ versus the mass within $5 r_e$ as measured by the GC velocities \citep{alabi_etal2017} and then use an NFW model to extrapolate
$M(<5 r_e)$ out to $r_{200}$.  From this, they find $M_{GCS} \propto M(r_{200})^{0.99}$ over the mass range
$M_{200} = 10^{12-14} M_{\odot}$, quite consistent with our result and with the previous literature that extends the same relation to much lower mass. 

The evidence for the scaling of $M_{GCS}$ is consistent with previous suggestions that the GCS mass was largely unaffected by feedback.  Once the massive, dense globular clusters form, they can be eroded by dynamical evolution within the galaxy, particularly tidal stripping, but will avoid the feedback effects that were more damaging to the star-forming gas within the galaxy. The ability of the X-ray gas to absorb feedback energy and heat up may be responsible for its higher 
specific energy than either the blue or red GCs \citep{forbes_etal2012}.

The physical meaning of the direct proportionality of $M_{GCS}$ to halo mass has been discussed by, for example, \cite{blakeslee1999},  HHH1, HBH,  \cite{choksi_etal2018}, \cite{choksi_gnedin2019}, and \cite{kruijssen2015}, with the suggestion that GC formation took place in roughly direct proportion to the gas mass originally present in the dark-matter halos that later assembled into bigger galaxies.  Other recent studies \citep{forbes_etal2018b, lim_etal2018, prole_etal2019} have pushed the relation down into the dwarf regime (as low as $M_h \simeq 10^9 M_{\odot}$), where increased scatter about the direct proportionality becomes a dominant feature.  An important new addition to modelling of the correlation is a more statistically based explanation \citep{choksi_etal2018,el-badry_etal2019,burkert_forbes2019}:  assuming that the small seed halos at the base of hierarchical merging form an initial GC population, even if there is a large variety of GC numbers among these small halos, the action of merging and the central limit theorem will lead to the observed direct proportionality for larger galaxies \citep[see also earlier comments in][]{kravtsov_gnedin2005,hhh2015}.  One \emph{caveat} to this effect has to do with the metal-richer GCs, which are essentially not present in the initial dwarf halos and arise in later stages of merging in more massive gas-rich halos.  The mass in these red GCs increases more steeply as $M_{GCS}(red) \sim M_h^{1.2}$ \citep{hhh2015}.  A more comprehensive explanation is needed to account for both the combination of halos with existing (mostly blue) GCs and the formation of later GCs along the merger tree \citep{choksi_gnedin2019,pfeffer_etal2018,el-badry_etal2019}.

The near-linear scaling of $M_X$ with $M_h$ is intriguing and may have a slightly more complex origin.  B18 (see their Fig.~7) find empirically that $M_X \propto M_{grav}^{1.7 \pm 0.2}$, a visible effect of the increasing proportion of high-temperature gas in progressively more massive galaxies.  When combined with our correlation $M_X \propto M_h^{1.0}$ from a smaller sample of galaxies (Fig.~2 middle panel), this gives
$M_{grav} \propto M_h^{0.5 \pm 0.1}$, which agrees with the trend shown in the bottom panel of Fig.~2.  The message we read from the correlations in Fig.~2 is that with increasing galaxy mass, the \emph{mass fraction} of X-ray gas increases.  But the fraction of dark matter \emph{also} increases at almost the same rate, and the combined effect is to keep the total mass of high-temperature gas nearly proportional to total halo mass.  

The observed trend $M_{grav} \propto M_h^{0.5}$ is well defined from our dataset, limited though it is.
$M_{grav}$ consists of baryonic matter (stars plus gas) plus the dark matter within $5 r_e$, and the stars plus gas make up a significant fraction of this \citep[see also][]{alabi_etal2017}. By contrast, dark matter makes up the great majority of the total mass $M_h$ in these large galaxies, and the mass fraction of dark matter keeps increasing with increasing galaxy luminosity, leading to the shallow dependence of $M_{grav}$ with $M_h$ that is observed.

\begin{table*}[t!]
\caption{Derived Slopes for Correlations}\label{tab2}
\centering\begin{tabular}{lccccc}
\hline\\
{Correlation} &{Slope} & Coef. & Intr. Scatter \\
&&\\
\hline\\
log M$_{GCS}$ vs. log $T_X$ & 1.8$\pm$0.2  & 0.80 & 0.13$\pm$0.03\\
log M$_{GCS}$ vs. log $L_X$ & 0.5$\pm$0.06  & 0.78 & 0.14$\pm$0.03\\
log M$_{GCS}$ vs. log $M_h$ & 1.03$\pm$0.03  & 0.90 & 0.11$\pm$0.03 \\
log $M_X$ vs. log $M_h$ & 1.0$\pm$0.2  & 0.65 & 0.41$\pm$0.10  \\
log M$_{grav}$ vs. log $M_h$ & 0.5$\pm$0.1  & 0.67 & 0.08$\pm$0.02\\
log M$_{(\star +X)}$ vs. log $M_{h}$ & 0.6$\pm$0.04  & 0.91 & 0.02$\pm$0.005 \\
log M$_{GCS}$/M$_X$ vs. log $M_h$ & -0.1$\pm$0.2 & -0.2 & 0.37$\pm$0.09\\

&\\
\hline\\
%log M$_{grav}$/M$_h$ vs. log $M_h$ & -0.6$\pm$0.05 & log() & -0.77 & \\
\end{tabular}
\end{table*}

It is perhaps worth mentioning that the central supermassive black hole (SMBH) mass has been shown to correlate almost one-to-one with $M_{GCS}$ and $M_h$
\citep{spitler_forbes2009,burkert_tremaine2010,gharris_etal2014}.  Though causal physical links between the
SMBH and globular cluster system seem unlikely \citep[see][]{gharris_etal2014}, the link between the SMBH 
and the X-ray gas is more direct (through AGN feedback).  Thus to first order, we have empirically 
determined scalings among three prominent components of the galaxy whose origins are at high redshift:
$M_{GCS} \sim M_X \sim M_{SMBH} \sim M_h$.

\section{Dark Matter Fraction Within $5 \lowercase{r_e}$}

The mass profiles of ETGs should be increasingly dominated by dark matter at larger
radii.  In the inner regions ($r \lesssim r_e$), important details of the evolution of a
galaxy including the epochs and amounts of gas infall and merging should strongly affect
the relative amounts of dark versus baryonic matter.  Contemporary cosmological hydrodynamic models \citep[e.g.][]{remus+2017,lovell+2018} predict that within $1 r_e$ the ratio of DM to
total mass $f_{DM}$ ranges over $\sim 0.2$ to 0.7 for large galaxies and is also a strong
function of redshift.  A higher \emph{in situ} fraction of stellar mass should go along
with lower $f_{DM}$, while a larger fraction of dry merging (accretion)
should extend the stellar mass profile and increase the inner $f_{DM}$.

\begin{figure}
    \centering
    \includegraphics[width=0.9\textwidth]{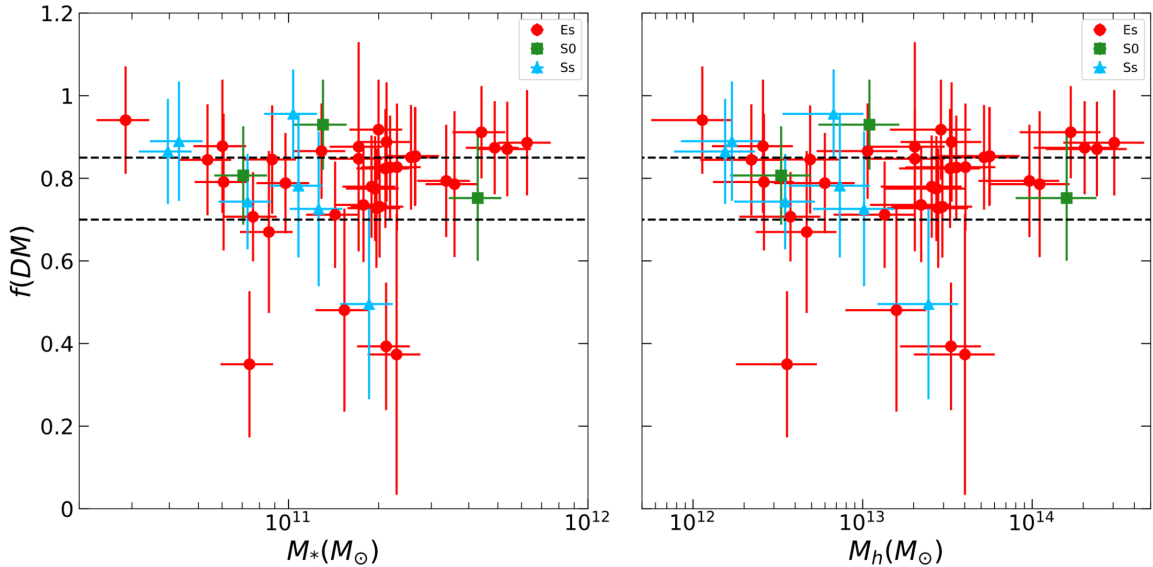}
    \vspace{-12.8cm}
    \caption{The dark matter fraction $f_{DM}$ within $5r_e$ is plotted versus
    $M_{\star}$ (left panel) or $M_h$ (right panel). The pair of horizontal
    \emph{dashed lines} at 0.70 and 0.85 give the upper and lower boundaries
    enclosing the great majority of models by \citet{remus+2017}, as quoted
    by \citet{alabi_etal2017}.}
    \label{fig:fDM}
\end{figure}

At much larger radii, however,
a more nearly uniform DM fraction might be expected \citep[cf.][]{deason_etal2012,wojtak_mamon2013,lovell+2018}. 
Recent discussions have focussed on the DM fraction within fiducial radii of either $1 r_e$ or $5 r_e$,
where the latter is large
enough to avoid much of the dispersion seen in the core
region.  Our data can be used to give a new observational assessment of $f_{DM}$, since $M_{grav}$ 
as used here (Eq.~(4)) directly measures the total gravitating mass within $5r_e$.  
We can then define
\begin{equation}
    f_{DM} = 1 - {(M_{\star} + M_X) \over M_{grav} }
\end{equation}
and plot this ratio versus either $M_{\star}$ or $M_h$.  Both versions are
shown in Figure \ref{fig:fDM}.
(Note that here, $M_{\star}$ is the total stellar mass of the galaxy rather than
the mass within $5r_e$.  However, for a S\'ersic/deVaucouleurs profile with n=4
typical for early-type galaxies, 90\% of the total light lies within $5r_e$, so
this is only a second-order correction.  The same issue applies to $M_X$, but
the X-ray gas mass is usually only a few percent of $M_{\star}$,
so its correction would be even smaller. The net effect is to make our $f_{DM}$ values
too large by about 2 percent on average, which we ignore for the present.)

Of the 45 galaxies in our sample, 40 fall within a narrow band with mean 
$\langle f_{DM} \rangle = 0.83$ and rms dispersion $\sigma_f = \pm 0.07$.
Little trend is seen with host galaxy mass, either versus $M_{\star}$ or $M_h$.
These results present a more uniform pattern for $f_{DM}$ versus galaxy mass
than has previously been thought to be the case \citep[e.g.][]{napolitano+2009,alabi_etal2017}
\citep[though see][]{wojtak_mamon2013}.
Various analytical models for galaxy mass profiles built from standard parameters 
for the DM halo profile shape, the stellar-to-halo mass
ratio versus galaxy mass, and the effective radii and central concentrations of the
DM and stellar profiles, indicate that $f_{DM}$ should gradually increase from $\sim 0.6$
at $M_{\star} \sim 10^{10} M_{\odot}$ up to 0.9 and above for the most massive 
galaxies \citep[e.g.][]{napolitano+2009,deason_etal2012,alabi_etal2016}.
However, recent models built on cosmological hydrodynamic simulations
\citep{wu+2014,remus+2017,lovell+2018} notably predict shallower trends with mass for $f_{DM}$
at the fiducial $5 r_e$.
The range of expected values from the Magneticum simulations \citep{remus+2017} in particular is indicated
by the dashed lines in Fig.~\ref{fig:fDM}; some individual cases scatter down to $\simeq 0.6$,
but the great majority fall in the interval $0.70 - 0.85$.\footnote{The Remus 
et al.~predictions for $5r_e$ are not
given directly in their 2017 paper, but can be seen in Figure 2 of \citet{alabi_etal2017}.} 
The agreement of these models with the measurements is striking.  Our results are also 
very much in the range found from the IllustrisTNG simulations \citep{lovell+2018}, and
in a large sample of isolated SDSS galaxies from satellite kinematics \citep{wojtak_mamon2013},
which both yield $f_{DM} \simeq 0.75 - 0.9$ for galaxies in the mass range that we discuss here.

The observational correlation of $f_{DM}$ with $M_{\star}$ has recently been analyzed 
by \citet{alabi_etal2016,alabi_etal2017} (hereafter A16, A17).  In their studies  
satellite kinematics including GCs and PNe are used to derive
the galaxy mass profiles and also the total mass within $5 r_e$.  
In their results, the individual galaxies scatter across all values from 
$f_{DM} \sim 0.1$ up to more than 0.9.  Although most of their program galaxies
fall near $f_{DM} \sim 0.8$, many scatter to much lower values 
and (as in our study) no clear systematic trend is seen with total stellar mass
(see particularly Fig.~2 of A17).  

There are two notable differences between our method and A16, A17 for determining $f_{DM}$.
\begin{enumerate}
    \item A16, A17 do not include gas mass in their calculation, which means that our values for $f_{DM}$ will be systematically lower than theirs for the same $M_{\star}$ and $M_{grav}$.
    However, $M_{\star}$ usually dominates the baryonic mass and so this 
    does not generate a major difference:  in most cases (though not all), $M_X$ is 
    only a few percent of $M_{\star}$.  If we were to neglect $M_X$ and recalculate our
    $f_{DM}$ values as $f_{DM} = 1 - (M_{\star}/M_{grav})$, we find a mean value 
    $\langle f_{DM} \rangle = 0.84$ with a dispersion of $\pm 0.08$.
    \item More importantly, we calculate $M_{grav}$ (the mass within $5 r_e$) in a completely different way.  Here, $M_{grav}$ (see Eq. (4))
    depends only on the density gradient and temperature of the X-ray gas, and the basic
    assumption of hydrostatic equilibrium.  A16, A17 use the kinematics of satellites
    along with the Tracer Mass Estimator technique.  Significant and well known 
    uncertainties in this method and in previous methods using satellite tracers 
    include the anisotropy of the GC orbital distribution, the slope of
    the gravitational potential, and the presence of substructure, all of which can
    differ strongly and unpredictably from one target galaxy to another.  
    In addition, most GCs (like the
    stellar light) lie at radii well within $5 r_e$ and so 
    the result for $M(< 5 r_e)$ for any one galaxy may depend heavily on small numbers of GCs that
    lie at the largest observed radii.  These issues and others are discussed at length 
    by A16, A17.
\end{enumerate}

\begin{figure}
    \centering
    \vspace{-1.5cm}
    \includegraphics[width=1.0\textwidth]{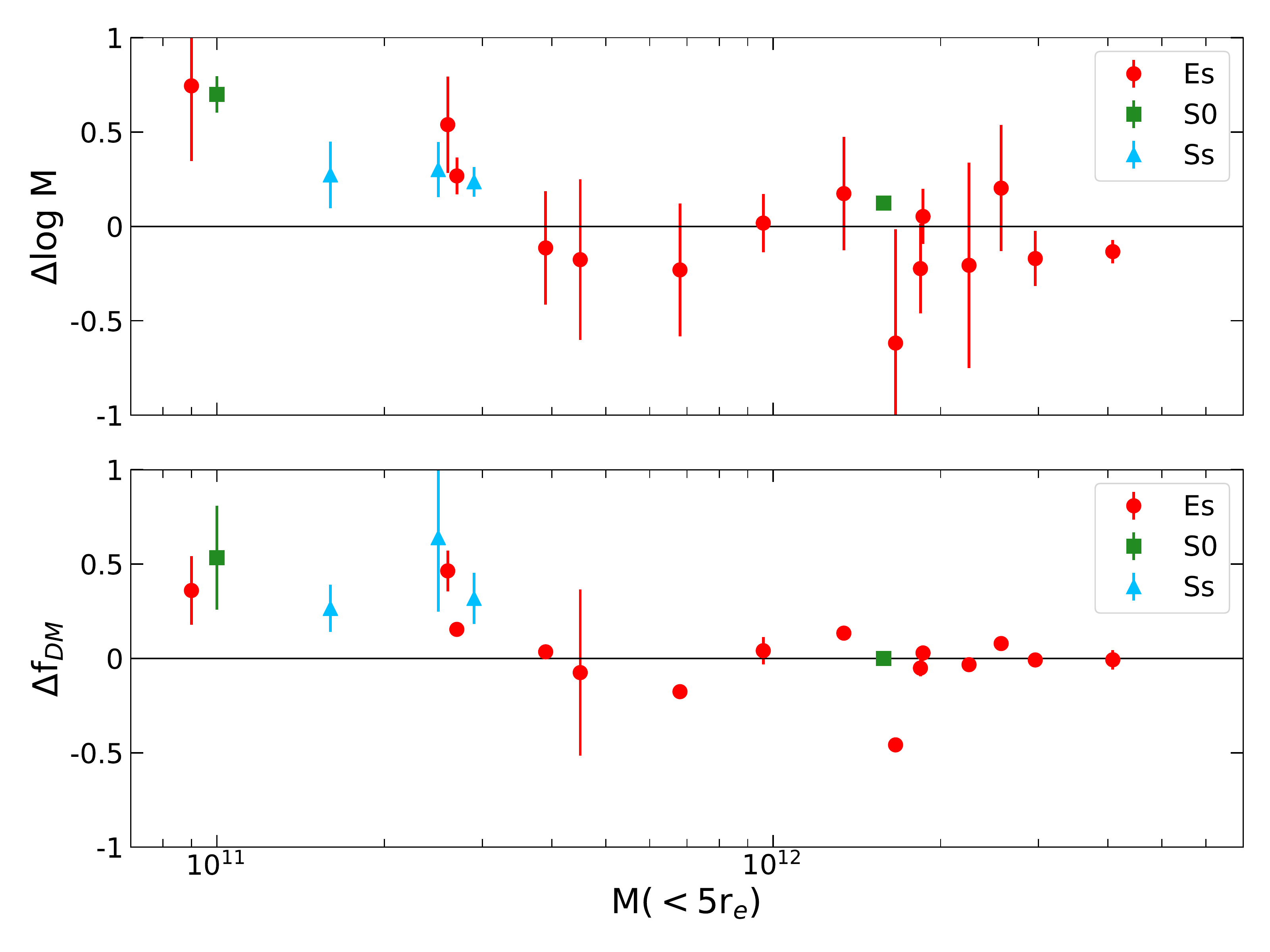}
    \vspace{-0.2cm}
    \caption{\emph{Upper panel:} $\Delta M_{grav}$ from our X-ray data is plotted against
    $M(5 r_e)$ from A17, For the 20 galaxies in common between this study and
    \citet{alabi_etal2017}, the logarithmic difference in mass measured within $5 r_e$ is
    plotted versus mass, as described in the text.
    \emph{Lower panel:} Difference $\Delta f_{DM} = f_{DM} - f_{DM}(A17)$ is plotted 
    versus $M(<5 r_e)$.  Here, our $f_{DM}$ values include only $M_{star}$ and not $M_X$,
    to make them more strictly comparable with the values from A17.}
    \vspace{0.5cm}
    \label{fig:2DM}
\end{figure}

Between the A17 dataset and ours there are 20
galaxies in common.   Key comparisons for these are shown in Figure \ref{fig:2DM}.
In the upper panel, the mass difference between $M_{grav}$ (this study) and
$M(<5r_e)$ (from A17) is plotted versus $M(<5r_e)$, where we have taken the masses
from A17 assuming $\beta=0$ for the satellite anisotropy parameter.
If the X-ray technique and the satellite kinematics technique are
both systematically valid, then they should simply scatter around $\Delta {\textrm log} M = 0$.  
The comparison shows good overall agreement
between the two mass measurement techniques for the higher-mass systems 
(log $M(<5r_e) \gtrsim 11.5$), but at lower mass, the (admittedly small number of)
datapoints indicate that $M_{grav}$ is systematically 
higher. The two sets of $f_{DM}$ values reveal a similar trend.  In the lower
panel of Fig.~\ref{fig:2DM}, the difference 
$\Delta f_{DM} = f_{DM}$(this study)$- f_{DM}$(A17) 
is shown, where now the values from our study include only
$M_{\star}$ and not the gas mass $M_X$, to make the two datasets more strictly
comparable.  For the higher-mass galaxies, good agreement is seen, but for the
lower-mass systems our $f_{DM}$ values tend to be $\simeq 0.3$ higher.  The main
reason for this offset is that $M_{grav}$ tends to be higher than
$M(<5r_e)$ in that range by factors of 2 or more.   

Possible reasons for the offset at low mass that may originate from the treatment of
the X-ray density and temperature distributions have already been mentioned above
in Section 3.  It is not yet clear, however, whether the uncertainties connected
with the X-ray analysis would produce a systematic error as opposed to a mere increase in
scatter.  In addition to the agreement with recent models (Fig. \ref{fig:fDM} and the discussion above), it is
worth noting that no suggestion of an offset at the lower mass end is seen in the correlations of $M_{grav}$ with
the other quantities in 
Figs. 2 and 3.  The comparisons shown in Fig.~\ref{fig:2DM} are, however, 
still based on few points, and a much larger overlap sample would be valuable.

In our list of measured $f_{DM}$ values, five galaxies 
(NGC 2768, 4365, 4382, 4697, and 5813) fall below $f_{DM} = 0.5$.
Low DM fractions could result from a variety of reasons, including
recent major accretion of gas and star formation activity in the inner halo   
\citep[e.g.][]{deason_etal2012,remus+2017}; or more prosaically 
undetected errors in $M_{grav}$ or $M_X$.
In these cases, however, we believe the lower values may be real.
For three of them (NGC 2768, 4697, 5813), the $M_X$ values are unusually large,
and excluding the gas from the calculation of $f_{DM}$ as above would raise the
DM fractions of all three above 0.6 and move them into the main spread of datapoints
in Fig.~\ref{fig:fDM}.  These three also show evidence for a history of gas-rich mergers
\citep{forbes+2016,crocker+2008,spiniello+2015,randall+2015}.
NGC 4382 is a recent major merger remnant \citep{ko+2018}, as is NGC 4365
\citep[e.g.][]{forbes+2016}, though neither of them now hold excessively large amounts
of X-ray gas.

\section{Summary}

In this study we have used recent measurements of globular cluster systems in massive galaxies,
and the properties of their X-ray gaseous atmospheres, to intercompare these two very different
types of data.  We find that the total GCS mass, the total mass in the X-ray gas, and the
total gravitating mass within $5 r_e$ can all be described accurately as simple power laws
versus each other or versus the total halo mass (essentially $M_{200}$) of the galaxy,
although with different slopes.  We find scalings $M_{GCS} \sim M_X \sim M_h^{1.0}$
and $M_{grav} = M(<5r_e) \sim M_h^{0.5}$.  Thus the mass ratio $(M_{GCS}/M_X)$ stays nearly 
constant with halo mass.

Our data also allow a new assessment of the mass fraction $f_{DM}$ of dark matter
within the fiducial radius $5 r_e$, based on a mass measurement technique independent of the
more normally used satellite kinematics methods.  Over the mass range of our sample
($10^{10} M_{\odot} \lesssim M_{\star} \lesssim 10^{12} M_{\odot}$), we find that $f_{DM}$ stays nearly
uniform with a mean at 0.83 and rms scatter $\pm 0.07$, and with only a few outliers at
lower values. This pattern differs from some other observational studies but generally 
agrees well with predictions from two recent suites of
cosmological hydrodynamic models.  Uncertainties remain, however,
about the accuracy of the X-ray-based mass measurements at the low-mass end that will need
to be explored further.

\section*{Acknowledgements}
This research has made use of data obtained from the Chandra Data Archive and the Chandra Source Catalog, and software provided by the Chandra X-ray Center (CXC) in the application packages CIAO, ChIPS, and Sherpa. We thank all the staff members involved in the Chandra project. Additionally, we have used ADS facilities.  We thank Mike Hudson for helpful discussions.  WEH and BRM acknowledge the financial support of NSERC.
%\makeatletter\@chicagotrue\makeatother

%\bibliographystyle{apj}
\bibliography{xray}

\begin{thebibliography}{}
\expandafter\ifx\csname natexlab\endcsname\relax\def\natexlab#1{#1}\fi
\providecommand{\url}[1]{\href{#1}{#1}}

\bibitem[{{Agertz} \& {Kravtsov}(2016)}]{agertz_kravtsov2016}
{Agertz}, O., \& {Kravtsov}, A.~V. 2016, \apj, 824, 79

\bibitem[{{Alabi} {et~al.}(2016){Alabi}, {Forbes}, {Romanowsky}, {Brodie},
  {Strader}, {Janz}, {Pota}, {Pastorello}, {Usher}, {Spitler}, {Foster},
  {Jennings}, {Villaume}, \& {Kartha}}]{alabi_etal2016}
{Alabi}, A.~B., {Forbes}, D.~A., {Romanowsky}, A.~J., {et~al.} 2016, \mnras,
  460, 3838

\bibitem[{{Alabi} {et~al.}(2017){Alabi}, {Forbes}, {Romanowsky}, {Brodie},
  {Strader}, {Janz}, {Usher}, {Spitler}, {Bellstedt}, \&
  {Ferr{\'e}-Mateu}}]{alabi_etal2017}
---. 2017, \mnras, 468, 3949

\bibitem[{{Arnaud}(1996)}]{arn}
{Arnaud}, K.~A. 1996, in Astronomical Society of the Pacific Conference Series,
  Vol. 101, Astronomical Data Analysis Software and Systems V, ed. G.~H.
  {Jacoby} \& J.~{Barnes}, 17

\bibitem[{{Babyk} {et~al.}(2018){Babyk}, {McNamara}, {Nulsen}, {Hogan},
  {Vantyghem}, {Russell}, {Pulido}, \& {Edge}}]{babyk_etal2018}
{Babyk}, I.~V., {McNamara}, B.~R., {Nulsen}, P.~E.~J., {et~al.} 2018, \apj,
  857, 32

\bibitem[{{Behroozi} {et~al.}(2013){Behroozi}, {Wechsler}, \&
  {Conroy}}]{behroozi_etal2013}
{Behroozi}, P.~S., {Wechsler}, R.~H., \& {Conroy}, C. 2013, \apj, 770, 57

\bibitem[{{Bell} {et~al.}(2003){Bell}, {McIntosh}, {Katz}, \&
  {Weinberg}}]{bell_etal2003}
{Bell}, E.~F., {McIntosh}, D.~H., {Katz}, N., \& {Weinberg}, M.~D. 2003, \apjs,
  149, 289

\bibitem[{{Blakeslee}(1997)}]{blakeslee1997}
{Blakeslee}, J.~P. 1997, \apjl, 481, L59

\bibitem[{{Blakeslee}(1999)}]{blakeslee1999}
---. 1999, \aj, 118, 1506

\bibitem[{{Boroson} {et~al.}(2011){Boroson}, {Kim}, \& {Fabbiano}}]{boroson}
{Boroson}, B., {Kim}, D.-W., \& {Fabbiano}, G. 2011, \apj, 729, 12

\bibitem[{{Brown} {et~al.}(2018){Brown}, {Casertano}, {Strader}, {Riess},
  {VandenBerg}, {Soderblom}, {Kalirai}, \& {Salinas}}]{brown_etal2018}
{Brown}, T.~M., {Casertano}, S., {Strader}, J., {et~al.} 2018, \apjl, 856, L6

\bibitem[{{Burkert} \& {Forbes}(2019)}]{burkert_forbes2019}
{Burkert}, A., \& {Forbes}, D. 2019, arXiv e-prints, arXiv:1901.00900

\bibitem[{{Burkert} \& {Tremaine}(2010)}]{burkert_tremaine2010}
{Burkert}, A., \& {Tremaine}, S. 2010, \apj, 720, 516

\bibitem[{{Cavaliere} \& {Fusco-Femiano}(1978)}]{caval}
{Cavaliere}, A., \& {Fusco-Femiano}, R. 1978, \aap, 70, 677

\bibitem[{{Choksi} \& {Gnedin}(2019)}]{choksi_gnedin2019}
{Choksi}, N., \& {Gnedin}, O.~Y. 2019, arXiv e-prints, arXiv:1905.05199

\bibitem[{{Choksi} {et~al.}(2018){Choksi}, {Gnedin}, \& {Li}}]{choksi_etal2018}
{Choksi}, N., {Gnedin}, O.~Y., \& {Li}, H. 2018, \mnras, 480, 2343

\bibitem[{{Crocker} {et~al.}(2008){Crocker}, {Bureau}, {Young}, \&
  {Combes}}]{crocker+2008}
{Crocker}, A.~F., {Bureau}, M., {Young}, L.~M., \& {Combes}, F. 2008, \mnras,
  386, 1811

\bibitem[{{Deason} {et~al.}(2012){Deason}, {Belokurov}, {Evans}, \&
  {McCarthy}}]{deason_etal2012}
{Deason}, A.~J., {Belokurov}, V., {Evans}, N.~W., \& {McCarthy}, I.~G. 2012,
  \apj, 748, 2

\bibitem[{{El-Badry} {et~al.}(2019){El-Badry}, {Quataert}, {Weisz}, {Choksi},
  \& {Boylan-Kolchin}}]{el-badry_etal2019}
{El-Badry}, K., {Quataert}, E., {Weisz}, D.~R., {Choksi}, N., \&
  {Boylan-Kolchin}, M. 2019, \mnras, 482, 4528

\bibitem[{{Ettori}(2000)}]{ettori}
{Ettori}, S. 2000, \mnras, 318, 1041

\bibitem[{{Fabjan} {et~al.}(2011){Fabjan}, {Borgani}, {Rasia}, {Bonafede},
  {Dolag}, {Murante}, \& {Tornatore}}]{Fabjan:11}
{Fabjan}, D., {Borgani}, S., {Rasia}, E., {et~al.} 2011, \mnras, 416, 801

\bibitem[{{Forbes} {et~al.}(2012){Forbes}, {Ponman}, \&
  {O'Sullivan}}]{forbes_etal2012}
{Forbes}, D.~A., {Ponman}, T., \& {O'Sullivan}, E. 2012, \mnras, 425, 66

\bibitem[{{Forbes} {et~al.}(2018{\natexlab{a}}){Forbes}, {Read}, {Gieles}, \&
  {Collins}}]{forbes_etal2018b}
{Forbes}, D.~A., {Read}, J.~I., {Gieles}, M., \& {Collins}, M.~L.~M.
  2018{\natexlab{a}}, \mnras, 481, 5592

\bibitem[{{Forbes} {et~al.}(2016){Forbes}, {Romanowsky}, {Pastorello},
  {Foster}, {Brodie}, {Strader}, {Usher}, \& {Pota}}]{forbes+2016}
{Forbes}, D.~A., {Romanowsky}, A.~J., {Pastorello}, N., {et~al.} 2016, \mnras,
  457, 1242

\bibitem[{{Forbes} {et~al.}(2018{\natexlab{b}}){Forbes}, {Bastian}, {Gieles},
  {Crain}, {Kruijssen}, {Larsen}, {Ploeckinger}, {Agertz}, {Trenti},
  {Ferguson}, {Pfeffer}, \& {Gnedin}}]{forbes_etal2018a}
{Forbes}, D.~A., {Bastian}, N., {Gieles}, M., {et~al.} 2018{\natexlab{b}},
  Proceedings of the Royal Society of London Series A, 474, 20170616

\bibitem[{{Goulding} {et~al.}(2016){Goulding}, {Greene}, {Ma}, {Veale},
  {Bogdan}, {Nyland}, {Blakeslee}, {McConnell}, \& {Thomas}}]{gould}
{Goulding}, A.~D., {Greene}, J.~E., {Ma}, C.-P., {et~al.} 2016, \apj, 826, 167

\bibitem[{{Harris} {et~al.}(2014){Harris}, {Poole}, \&
  {Harris}}]{gharris_etal2014}
{Harris}, G.~L.~H., {Poole}, G.~B., \& {Harris}, W.~E. 2014, \mnras, 438, 2117

\bibitem[{{Harris}(2016)}]{harris2016}
{Harris}, W.~E. 2016, \aj, 151, 102

\bibitem[{{Harris} {et~al.}(2017{\natexlab{a}}){Harris}, {Blakeslee}, \&
  {Harris}}]{hbh2017}
{Harris}, W.~E., {Blakeslee}, J.~P., \& {Harris}, G.~L.~H. 2017{\natexlab{a}},
  \apj, 836, 67

\bibitem[{{Harris} {et~al.}(2017{\natexlab{b}}){Harris}, {Ciccone}, {Eadie},
  {Gnedin}, {Geisler}, {Rothberg}, \& {Bailin}}]{harris_etal2017}
{Harris}, W.~E., {Ciccone}, S.~M., {Eadie}, G.~M., {et~al.} 2017{\natexlab{b}},
  \apj, 835, 101

\bibitem[{{Harris} {et~al.}(2015){Harris}, {Harris}, \& {Hudson}}]{hhh2015}
{Harris}, W.~E., {Harris}, G.~L., \& {Hudson}, M.~J. 2015, \apj, 806, 36

\bibitem[{{Harris} {et~al.}(2013){Harris}, {Harris}, \& {Alessi}}]{hha2013}
{Harris}, W.~E., {Harris}, G.~L.~H., \& {Alessi}, M. 2013, \apj, 772, 82

\bibitem[{{Hudson} {et~al.}(2014){Hudson}, {Harris}, \& {Harris}}]{hhh2014}
{Hudson}, M.~J., {Harris}, G.~L., \& {Harris}, W.~E. 2014, \apjl, 787, L5

\bibitem[{{Hudson} {et~al.}(2015){Hudson}, {Gillis}, {Coupon}, {Hildebrandt},
  {Erben}, {Heymans}, {Hoekstra}, {Kitching}, {Mellier}, {Miller}, {Van
  Waerbeke}, {Bonnett}, {Fu}, {Kuijken}, {Rowe}, {Schrabback}, {Semboloni},
  {van Uitert}, \& {Velander}}]{hudson_etal2015}
{Hudson}, M.~J., {Gillis}, B.~R., {Coupon}, J., {et~al.} 2015, \mnras, 447, 298

\bibitem[{{Kelly}(2007)}]{Kelly:07}
{Kelly}, B.~C. 2007, \apj, 665, 1489

\bibitem[{{Kim} {et~al.}(2019){Kim}, {James}, {Fabbiano}, {Forbes}, \&
  {Alabi}}]{james_etal2019}
{Kim}, D.-W., {James}, N., {Fabbiano}, G., {Forbes}, D., \& {Alabi}, A. 2019,
  \mnras, 488, 1072

\bibitem[{{Ko} {et~al.}(2018){Ko}, {Lee}, {Park}, {Sohn}, {Lim}, \&
  {Hwang}}]{ko+2018}
{Ko}, Y., {Lee}, M.~G., {Park}, H.~S., {et~al.} 2018, \apj, 859, 108

\bibitem[{{Kravtsov} \& {Gnedin}(2005)}]{kravtsov_gnedin2005}
{Kravtsov}, A.~V., \& {Gnedin}, O.~Y. 2005, \apj, 623, 650

\bibitem[{{Kravtsov} {et~al.}(2018){Kravtsov}, {Vikhlinin}, \&
  {Meshcheryakov}}]{kravtsov_etal2018}
{Kravtsov}, A.~V., {Vikhlinin}, A.~A., \& {Meshcheryakov}, A.~V. 2018,
  Astronomy Letters, 44, 8

\bibitem[{{Kruijssen}(2015)}]{kruijssen2015}
{Kruijssen}, J.~M.~D. 2015, \mnras, 454, 1658

\bibitem[{{Kruijssen} {et~al.}(2019){Kruijssen}, {Pfeffer}, {Crain}, \&
  {Bastian}}]{kruijssen_etal2019}
{Kruijssen}, J.~M.~D., {Pfeffer}, J.~L., {Crain}, R.~A., \& {Bastian}, N. 2019,
  \mnras, arXiv:1904.04261

\bibitem[{{Leaman} {et~al.}(2013){Leaman}, {VandenBerg}, \&
  {Mendel}}]{leaman_etal2013}
{Leaman}, R., {VandenBerg}, D.~A., \& {Mendel}, J.~T. 2013, \mnras, 436, 122

\bibitem[{{Leauthaud} {et~al.}(2012){Leauthaud}, {Tinker}, {Bundy}, {Behroozi},
  {Massey}, {Rhodes}, {George}, {Kneib}, {Benson}, {Wechsler}, {Busha},
  {Capak}, {Cort{\^e}s}, {Ilbert}, {Koekemoer}, {Le F{\`e}vre}, {Lilly},
  {McCracken}, {Salvato}, {Schrabback}, {Scoville}, {Smith}, \&
  {Taylor}}]{leauthaud_etal2012}
{Leauthaud}, A., {Tinker}, J., {Bundy}, K., {et~al.} 2012, \apj, 744, 159

\bibitem[{{Liedahl} {et~al.}(1995){Liedahl}, {Osterheld}, \&
  {Goldstein}}]{Liedahl}
{Liedahl}, D.~A., {Osterheld}, A.~L., \& {Goldstein}, W.~H. 1995, \apjl, 438,
  L115

\bibitem[{{Lim} {et~al.}(2018){Lim}, {Peng}, {C{\^o}t{\'e}}, {Sales}, {den
  Brok}, {Blakeslee}, \& {Guhathakurta}}]{lim_etal2018}
{Lim}, S., {Peng}, E.~W., {C{\^o}t{\'e}}, P., {et~al.} 2018, \apj, 862, 82

\bibitem[{{Lovell} {et~al.}(2018){Lovell}, {Pillepich}, {Genel}, {Nelson},
  {Springel}, {Pakmor}, {Marinacci}, {Weinberger}, {Torrey}, {Vogelsberger},
  {Alabi}, \& {Hernquist}}]{lovell+2018}
{Lovell}, M.~R., {Pillepich}, A., {Genel}, S., {et~al.} 2018, \mnras, 481, 1950

\bibitem[{{Mewe} {et~al.}(1986){Mewe}, {Lemen}, \& {van den Oord}}]{Mewe}
{Mewe}, R., {Lemen}, J.~R., \& {van den Oord}, G.~H.~J. 1986, \aaps, 65, 511

\bibitem[{{Mitchell} {et~al.}(2016){Mitchell}, {Lacey}, {Baugh}, \&
  {Cole}}]{mitchell_etal2016}
{Mitchell}, P.~D., {Lacey}, C.~G., {Baugh}, C.~M., \& {Cole}, S. 2016, \mnras,
  456, 1459

\bibitem[{{Moster} {et~al.}(2013){Moster}, {Naab}, \&
  {White}}]{moster_etal2013}
{Moster}, B.~P., {Naab}, T., \& {White}, S.~D.~M. 2013, \mnras, 428, 3121

\bibitem[{{Napolitano} {et~al.}(2009){Napolitano}, {Romanowsky}, {Coccato},
  {Capaccioli}, {Douglas}, {Noordermeer}, {Gerhard}, {Arnaboldi}, {de Lorenzi},
  {Kuijken}, {Merrifield}, {O'Sullivan}, {Cortesi}, {Das}, \&
  {Freeman}}]{napolitano+2009}
{Napolitano}, N.~R., {Romanowsky}, A.~J., {Coccato}, L., {et~al.} 2009, \mnras,
  393, 329

\bibitem[{{Pfeffer} {et~al.}(2018){Pfeffer}, {Kruijssen}, {Crain}, \&
  {Bastian}}]{pfeffer_etal2018}
{Pfeffer}, J., {Kruijssen}, J.~M.~D., {Crain}, R.~A., \& {Bastian}, N. 2018,
  \mnras, 475, 4309

\bibitem[{{Prole} {et~al.}(2019){Prole}, {Hilker}, {van der Burg}, {Cantiello},
  {Venhola}, {Iodice}, {van de Ven}, {Wittmann}, {Peletier}, {Mieske},
  {Capaccioli}, {Napolitano}, {Paolillo}, {Spavone}, \&
  {Valentijn}}]{prole_etal2019}
{Prole}, D.~J., {Hilker}, M., {van der Burg}, R.~F.~J., {et~al.} 2019, \mnras,
  484, 4865

\bibitem[{{Randall} {et~al.}(2015){Randall}, {Nulsen}, {Jones}, {Forman},
  {Bulbul}, {Clarke}, {Kraft}, {Blanton}, {David}, {Werner}, {Sun}, {Donahue},
  {Giacintucci}, \& {Simionescu}}]{randall+2015}
{Randall}, S.~W., {Nulsen}, P.~E.~J., {Jones}, C., {et~al.} 2015, \apj, 805,
  112

\bibitem[{{Remus} {et~al.}(2017){Remus}, {Dolag}, {Naab}, {Burkert},
  {Hirschmann}, {Hoffmann}, \& {Johansson}}]{remus+2017}
{Remus}, R.-S., {Dolag}, K., {Naab}, T., {et~al.} 2017, \mnras, 464, 3742

\bibitem[{{Spiniello} {et~al.}(2015){Spiniello}, {Napolitano}, {Coccato},
  {Pota}, {Romanowsky}, {Tortora}, {Covone}, \& {Capaccioli}}]{spiniello+2015}
{Spiniello}, C., {Napolitano}, N.~R., {Coccato}, L., {et~al.} 2015, \mnras,
  452, 99

\bibitem[{{Spitler} \& {Forbes}(2009)}]{spitler_forbes2009}
{Spitler}, L.~R., \& {Forbes}, D.~A. 2009, \mnras, 392, L1

\bibitem[{{Su} \& {Irwin}(2013)}]{su}
{Su}, Y., \& {Irwin}, J.~A. 2013, \apj, 766, 61

\bibitem[{{van Uitert} {et~al.}(2016){van Uitert}, {Cacciato}, {Hoekstra},
  {Brouwer}, {Sif{\'o}n}, {Viola}, {Baldry}, {Bland-Hawthorn}, {Brough},
  {Brown}, {Choi}, {Driver}, {Erben}, {Heymans}, {Hildebrandt}, {Joachimi},
  {Kuijken}, {Liske}, {Loveday}, {McFarland}, {Miller}, {Nakajima}, {Peacock},
  {Radovich}, {Robotham}, {Schneider}, {Sikkema}, {Taylor}, \& {Verdoes
  Kleijn}}]{vanUitert_etal2016}
{van Uitert}, E., {Cacciato}, M., {Hoekstra}, H., {et~al.} 2016, \mnras, 459,
  3251

\bibitem[{{Wechsler} \& {Tinker}(2018)}]{wechsler_tinker2018}
{Wechsler}, R.~H., \& {Tinker}, J.~L. 2018, \araa, 56, 435

\bibitem[{{Wojtak} \& {Mamon}(2013)}]{wojtak_mamon2013}
{Wojtak}, R., \& {Mamon}, G.~A. 2013, \mnras, 428, 2407

\bibitem[{{Wu} {et~al.}(2014){Wu}, {Gerhard}, {Naab}, {Oser},
  {Martinez-Valpuesta}, {Hilz}, {Churazov}, \& {Lyskova}}]{wu+2014}
{Wu}, X., {Gerhard}, O., {Naab}, T., {et~al.} 2014, \mnras, 438, 2701

\end{thebibliography}

\label{lastpage}

\end{document}